\documentclass[reprint,pra,amsmath,amssymb,aps,showpacs,preprintnumbers]{revtex4-1}
\usepackage{graphicx} 
\usepackage{datetime}
\usepackage{bm} 

\begin{document}
\title{Cooperative emission  mediated by cooperative energy transfer to a plasmonic antenna}
\author{Tigran V. Shahbazyan}
\affiliation{Department of Physics, Jackson State University, Jackson, Mississippi 39217 USA}


\begin{abstract} 
We develop  a theory of cooperative emission  mediated by cooperative energy transfer (CET)  from an ensemble of quantum emitters (QE) to plasmonic antenna at a rate equal to the sum of individual QE-plasmon energy transfer  rates. If the antenna radiation efficiency is sufficiently high, the transferred energy is radiated away at approximately the same cooperative rate that scales with the ensemble size. We derive explicit expressions, in terms of local fields, for cooperative Purcell factor and enhancement factor for power spectrum valid for plasmonic structures of any shape with characteristic size smaller than the radiation wavelength. The radiated power spectrum retains the plasmon resonance lineshape with overall amplitude scaling with the ensemble size. If QEs are located in a region with nearly constant plasmon local density of states (LDOS), e.g., inside a plasmonic nanocavity, we demonstrate that the CET  rate scales linearly with the number of excited QEs, consistent with the experiment, and can be tuned in a wide range by varying the excitation power. For QEs  distributed in an extended region saturating the plasmon mode volume, we show that the cooperative Purcell factor has universal form independent of the  system size. The CET mechanism   incorporates the plasmon LDOS enhancement  as well, giving rise to  possibilities of controlling  the emission rate beyond   field enhancement limits.

\end{abstract}

\maketitle


%
\section{Introduction}
\label{sec:intro}

Surface plasmons are collective electronic excitations that can be resonantly excited in metal-dielectric structures with characteristic size well below the diffraction limit \cite{maier-book}. Rapid oscillations of the electron charge density at the metal-dielectric interfaces generate extremely strong alternating local fields that can dramatically affect optical properties of nearby dye molecules or semiconductor quantum dots (QDs), hereafter referred to as quantum emitters (QEs) \cite{moskovits-rmp85,atwater-jap05,ozbay-science06,stockman-review}. Optical interactions between QEs and plasmons give rise to a number of major phenomena in plasmon-enhanced spectroscopy  such as surface-enhanced Raman scattering \cite{sers}, plasmon-enhanced fluorescence and luminescence \cite{novotny-prl06,sandoghdar-prl06,halas-nl07}, plasmon-assisted energy transfer (ET) \cite{lakowicz-jf03,andrew-science04,krenn-nl08,blum-prl12}, strong QE-plasmon coupling \cite{bellessa-prl04,sugawara-prl06,fofang-nl08,gomez-nl10,salomon-prl12,guebrou-prl12,antosiewicz-acsphotonics14} and plasmonic laser (spaser) \cite{bergman-prl03,stockman-natphot08,noginov-nature09}. 

A generic plasmonic effect underpinning many applications \cite{mikkelsen-np14,yablonovitch-pnas15,angelis-nl13,fiore-nn14,angelis-sr15,guo-np16} is the strong enhancement of spontaneous emission rate for an excited QE near a plasmonic structure occurring due to a highly efficient ET to a plasmon mode followed by plasmon's radiative decay (antenna effect) \cite{carminati-oc06,lalanne-prl13,pelton-np15,bonod-prb15,belov-sr15,greffet-acsph17,koenderink-acsphot17,lalanne-lpr18,shahbazyan-prb18}. High QE-plasmon ET rates are due to large plasmon density of states (LDOS) in small plasmonic systems which can dramatically exceed the corresponding density of photonic states distributed on a much larger spatial scale of radiation wavelength. In the regions of strong field confinement ("hot spots"), e.g., near sharp metal tips or in a gap between closely spaced metal structures, the  Purcell factor \cite{purcell-pr46}, which describes the decay rate enhancement relative to free-space decay, can reach several orders of magnitude. However, the local field enhancement of the decay rate  is ultimately limited  by losses  and nonlocal effects in metals \cite{smith-science12,mortensen-nc14}.

On the other hand, light  emission from an \textit{ensemble} of QEs can be greatly accelerated by cooperative effects arising from electromagnetic correlations between QEs. A prominent example of cooperative emission is the Dicke superradiance  of QEs interacting with the common radiation field which takes place at a rate scaling with the ensemble size \cite{dicke-pr54,haroche-pr82}.  Near a plasmonic structure,  the radiative coupling between QEs can be significantly enhanced due to resonant light scattering \cite{pustovit-prl09,pustovit-prb10,vidal-nl10,sanders-pra10,pustovit-prb13,may-prb14,greffet-pra16,lalanne-prb17}, which also reduces the  detrimental effect of the direct dipole interactions \cite{friedberg-pr73,haroche-pr82,shahbazyan-prb00}. At the same time, the Ohmic losses in metal suppress correlations between QEs close to the interface, where the local field enhancement is strongest \cite{pustovit-prl09,pustovit-prb10}, implying that plasmon-enhanced superradiance hinges on a delicate interplay between  direct dipole coupling,  plasmonic correlations, and Ohmic losses.

\begin{figure}[b]
\begin{center}
\includegraphics[width=0.85\columnwidth]{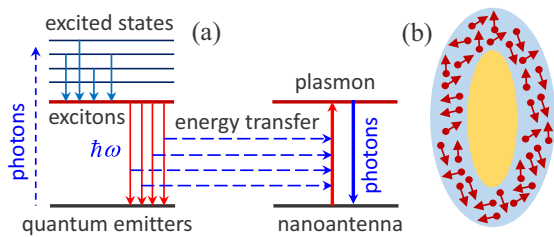}
\caption{\label{fig1} (a) Schematics of CET from QE ensemble to plasmonic antenna. (b) Schematics of plasmonic system with QEs distributed within dielectric shell surrounding metallic core. }
\end{center}
\end{figure}

In this paper, we describe another mechanism of cooperative emission based on \textit{cooperative energy transfer} (CET) from an ensemble of excited QEs to a plasmonic antenna  that is especially efficient in lossy plasmonic systems [see Fig.~\ref{fig1}(a)]. We demonstrate that  plasmonic correlations between QEs lead to the  emergence of a collective state that transfers its energy to  a resonant plasmon mode cooperatively, i.e., at a rate equal to the \textit{sum} of individual QE-plasmon ET rates. If the antenna's radiation efficiency is sufficiently high, a substantial part of this energy is radiated away at  approximately the same CET rate, which scales with the ensemble size, while the rest is mainly dissipated via the Ohmic losses in metal. Note that cooperative acceleration of the ensemble decay takes place \textit{on top} of  plasmonic enhancement of  individual QE decay rates and, therefore, the ensemble emission rate is conveniently described by \textit{cooperative Purcell factor}, which  incorporates both  the plasmon LDOS enhancement and the cooperative effects. At the same time,   the power spectrum radiated by the plasmonic antenna retains the  plasmon resonance lineshape  with overall amplitude scaling with the ensemble size. 

%
%
%
%

Importantly, the CET-based cooperative emission is a \textit{weak}  coupling effect that does not imply plasmon reabsorption, i.e., the energy transferred to a plasmon from a collective state, formed due to plasmonic correlations, is not transferred back to QEs but is either radiated away by the antenna or dissipated via the Ohmic losses in metal. This can take place if, e.g., the QE absorption maximum overlaps weakly with  plasmon's emission band, e.g., for dye molecules with a significant Stokes shift or in semiconductor QDs whose absorption maximum lies well above the emission line. As a result, the emission rate is proportional to the number of \textit{excited}, rather than total QEs and, therefore, can be tuned in a wide range by varying the excitation power. In contrast, in the strong coupling regime characterized by sustained QE-plasmon energy exchange, the QE-plasmon coupling scales with full QE ensemble size \cite{bellessa-prl04,sugawara-prl06,fofang-nl08,gomez-nl10,salomon-prl12,guebrou-prl12,antosiewicz-acsphotonics14}. Furthermore, since the ET rate between a donor (excited QE) and an acceptor (plasmon) is determined by  spectral overlap between the donor's emission and  the acceptor's absorption bands \cite{novotny-book}, the CET mechanism is largely insensitive to QE  frequency variations due to, e.g., direct dipole coupling or, in the case of QDs, their size variations, if they stay within the  broad plasmon absorption band. This stands in a stark contrast  to common cooperative mechanisms, such as superradiance, where the emission spectra are strongly affected by even   weak disorder in the QE frequencies \cite{friedberg-pr73,haroche-pr82,shahbazyan-prb00}.

The CET-based cooperative emission  was recently observed experimentally from CdSe/ZnS QDs placed inside a plasmonic nanocavity \cite{strangi-prl19}. In this experiment, the measured emission rate increased linearly with the excitation power implying that only excited QDs were involved in the cooperative emission. At the same time, the emission spectra exhibited the lineshape  of cavity plasmon resonance, but with overall amplitude scaling linearly with the excitation power. Both the above features  are characteristic for cooperative emission mediated by CET to plasmonic antenna, while the linear scaling indicates that QDs were located in a region with nearly constant plasmon LDOS, as we discuss below.  

We present a detailed discussion of another common experimental setup, in which the QEs are distributed within some region outside the plasmonic structure, e.g.,  within a dielectric shell enclosing a metal nanoparticle [see Fig.~\ref{fig1}(b)]. In this case, the plasmon LDOS and, accordingly, individual QE-plasmon ET rates fall  off rapidly away from the metal surface  and, therefore, the emission rate no longer scales linearly with the QE number  but, instead,   is determined by the \textit{average} plasmon LDOS in the QE region. If the QE region is sufficiently extended to \textit{saturate} the plasmon mode volume, i.e., the plasmon field spillover beyond the QE region is small, then the remote QEs do not, in fact, participate in the CET to  plasmonic antenna and so the emission rate saturates as well.  In this case, the cooperative Purcell factor takes a universal  (i.e., independent of LDOS) form for any metal-dielectric structure and depends on system geometry solely via the plasmon frequency  and its quality factor.  Finally, we present numerical results comparing cooperative acceleration of the ensemble decay rate to LDOS enhancement of the individual QE decay rate at a hot spot.

In our analysis, we rely on our plasmon Green's function approach \cite{shahbazyan-prl16} that we recently employed to study the emission by a single QE situated near a plasmonic antenna \cite{shahbazyan-prb18}. In this paper, we extend this approach to an ensemble of QEs near a plasmonic structure to describe the emergence of  collective states due to plasmonic correlations between the QEs. The collective state that interacts strongly with the plasmon mode is found explicitly  and its decay rate is shown  equal to the sum of individual QE rates. Within this approach, we also evaluate the full radiated power spectrum to find that it is dominated by the antenna, which radiates at a rate that scales with the ensembles size.

The paper is organized as follows. In Sec. \ref{sec:single}, we set up our approach by first treating a spontaneous decay of a single QE coupled to a plasmonic resonator. In Sec. \ref{sec:coop}, we present our theory of CET-based cooperative emission. In Sec. \ref{sec:num}, we discuss specific system geometries relevant for the experiment and present our numerical results. Section \ref{sec:conc} concludes the paper.

\section{Spontaneous decay of a quantum emitter near plasmonic resonator}
\label{sec:single}

We start with setting up our approach by first considering spontaneous decay of  a \textit{single} QE resonantly coupled to a plasmonic antenna, which represents a metal-dielectric nanostructure characterized by complex  dielectric function  $\varepsilon (\omega,\bm{r})=\varepsilon' (\omega,\bm{r})+i\varepsilon'' (\omega,\bm{r})$. 
The decay rate of a QE  located at $\bm{r}_{0}$  has the form \cite{novotny-book}
%
\begin{equation}
\gamma=\frac{2}{\hbar}\text{Im} [\bm{p}^{*} \!\cdot \bm{\mathcal{E}}(\omega,\bm{r}_{0})],
\end{equation}
where 
%
\begin{equation}
\bm{\mathcal{E}}(\omega,\bm{r} ) = \bar{\bm{D}}(\omega;\bm{r},\bm{r}_{0}) \bm{p}
\end{equation}
is  a slow component  of the electric field generated by the QE's oscillating dipole $\bm{p}e^{-i\omega t}=\mu\bm{n} e^{-i\omega t}$, where $\mu$ and  $\bm{n}$ are  the dipole moment and orientation, respectively, and 
%
\begin{equation}
\label{green-nf}
\bar{\bm{D}}(\omega;\bm{r},\bm{r}')=\frac{4\pi \omega^{2}}{c^{2}}\,\bar{\bm{G}}(\omega;\bm{r},\bm{r}')
\end{equation}
is the  electromagnetic dyadic Green's function (hereafter, a bar indicates a tensor in spatial indices). Here, $\bar{\bm{G}}$ is  the standard  Green's dyadic for   Maxwell equations that includes the plasmonic system and $c$ is speed of light. In the absence of a plasmonic system, the free-space Green's function $\bar{\bm{D}}_{0}(\omega;\bm{r},\bm{r}')=(4\pi \omega^{2}/c^{2})\,\bar{\bm{G}}_{0}(\omega;\bm{r},\bm{r}')$ yields the radiative decay rate of an isolated QE:  $\gamma_{0}^{r}=4\mu^{2}\omega^{3}/3\hbar c^{3}$.

\subsection{Plasmon Green's function}

We assume that the plasmonic system size is smaller than the radiation wavelength  and so the plasmon modes can be described within quasistatic approximation  \cite{stockman-review}  by the Gauss law $\bm{\nabla}\!\cdot\! \left [\varepsilon' (\omega_{m},\bm{r}) \bm{E}(\bm{r})\right ]=0$ with standard boundary conditions, where $\omega_{m}$ is the mode frequency  and $\bm{E}(\bm{r})$ is the mode electric field, which we chose to be real. 
In the near field limit,  the electromagnetic Green's function Eq.~(\ref{green-nf}) can be decomposed as  $\bar{\bm{D}}(\omega;\bm{r},\bm{r}')=\bar{\bm{D}}_{0}(\omega;\bm{r},\bm{r}')+\bar{\bm{D}}_{m}(\omega;\bm{r},\bm{r}')$, where  $\bar{\bm{D}}_{m}(\omega;\bm{r},\bm{r}')$ is the plasmon  Green's function. Here, we only consider the case  when the QE frequency $\omega$ is close to a particular mode frequency $\omega_{m}$ and therefore disregard contribution from the non-resonant modes (we will return to this point later). Near the resonance, the plasmon dyadic Green's function has the form \cite{shahbazyan-prl16,shahbazyan-prb18} 
\begin{equation}
\label{dyadic-mode}
\bar{\bm{D}}_{m}(\omega;\bm{r},\bm{r}') = \frac{\omega_{m}}{4 U_{m}}\frac{\bm{E}(\bm{r})\bm{E}(\bm{r}')}{\omega_{m}-\omega-i\gamma_{m}/2 },
\end{equation}
where $\gamma_{m}=W_{m}/U_{m}$ is  the plasmon decay rate, and the tensor product of electric fields is implied. Here, 
\begin{align}
\label{energy-mode}
U_{m}
= \frac{1}{16\pi} 
\!\int \!  dV \,   
\dfrac{\partial[\omega_{m}\varepsilon'(\omega_{m},\bm{r})]}{\partial \omega_{m}}\,\bm{E}^{2}(\bm{r})
\end{align}
is  the plasmon mode energy  and  $W_{m}=W_{m}^{\rm nr}+W_{m}^{r}$ is  the  dissipated power due to nonradiative ($W_{m}^{\rm nr}$) and radiative ($W_{m}^{r}$) losses. The former is primarily due to the Ohmic losses in metal and has the standard form \cite{landau} 
%
\begin{equation}
\label{power-nr}
W_{m}^{\rm nr}= \dfrac{\omega_{m}}{8\pi}\!\int \! dV  \varepsilon''(\omega_{m},\bm{r}) \bm{E}^{2}(\bm{r}).
\end{equation}
Since the integrations in $W_{m}^{\rm nr}$ and  $U_{m}$ are, in fact, restricted to the metallic regions with dielectric function $\varepsilon(\omega)=\varepsilon'(\omega)+i\varepsilon''(\omega)$,  the usual   expression for plasmon's nonradiative decay rate follows \cite{stockman-review}: 
%
\begin{equation}
\label{rate-nr}
\gamma_{m}^{\rm nr}=\frac{W_{m}^{\rm nr}}{U_{m}}=\frac{2\varepsilon''(\omega_{m})}{\partial \varepsilon'(\omega_{m})/\partial \omega_{m}}. 
\end{equation}
At the same time, if the system size is smaller than the radiation wavelength, the power \textit{ radiated} by the plasmonic antenna is similar to that of a  localized dipole \cite{novotny-book}, 
%
\begin{equation}
\label{power-rad}
W_{m}^{r}= \dfrac{\omega_{m}^{4}}{3c^{3}} \, \bm{\mathcal{P}}^{2},
\end{equation}
where $\bm{\mathcal{P}}=\!\int \! dV \chi'(\omega_{m},\bm{r})\bm{E}(\bm{r}) $ is the plasmon mode's dipole moment and $\chi(\omega,\bm{r})=[\varepsilon(\omega,\bm{r})-1]/4\pi$ is the plasmonic system's susceptibility. Note that, in this approximation, the real part of  susceptibility defines  the plasmon radiated power, while its imaginary part  determines the nonradiative losses Eq.~(\ref{power-nr}) \cite{shahbazyan-prb18}. Accordingly, the  radiative decay rate of a plasmon mode has the form 
%
\begin{equation}
\label{rate-rad}
\gamma_{m}^{r}=\frac{W_{m}^{r}}{U_{m}}=\dfrac{8\pi^{2}\omega_{m}}{3\lambda_{m}^{3}} \, \frac{\left [\int \! dV (\varepsilon'-1)\bm{E}\right ]^{2}}{\int \! dV \bm{E}^{2}\,\partial (\omega_{m}\varepsilon')/\partial \omega_{m}},
\end{equation}
where $\lambda_{m}=2\pi c/\omega_{m}$ is the plasmon wavelength and we denoted $\varepsilon\equiv \varepsilon(\omega_{m},\bm{r})$ in the integrand.  The rates Eqs.~(\ref{rate-nr}) and (\ref{rate-rad}) constitute the full plasmon decay rate $\gamma_{m}=\gamma_{m}^{\rm nr}+\gamma_{m}^{r}$. 
The plasmon radiation efficiency $\eta$ has the form
\begin{equation}
\label{mode-efficiency}
\eta=\frac{\gamma_{m}^{r}}{\gamma_{m}}=\dfrac{\zeta}{1+\zeta},
\end{equation}
where the parameter
\begin{equation}
\label{mode-rate-relative}
\zeta=\frac{\gamma_{m}^{r}}{\gamma_{m}^{\rm nr}}
=\frac{4\pi^{2}}{3 \lambda_{m}^{3}} 
\,
\frac{\left [ \int \! dV  (\varepsilon'-1)\bm{E}(\bm{r})\right ]^{2}}{ \int \! dV \varepsilon'' \bm{E}^{2}(\bm{r})},
\end{equation}
characterizes the relative magnitude of plasmon's radiative and nonradiative decay rates.  Note that, for small nanoplasmonic systems,  $\gamma_{m}^{\rm nr}$ should be amended to include the  surface-assisted Landau damping rate  \cite{shahbazyan-prb16}.  

\subsection{Purcell factor for spontaneous decay rate and radiated power spectrum}

The decay rate of a QE near a plasmonic system has the form $\gamma=\gamma_{0}^{r}+\gamma_{\rm et}$,  where 
\begin{equation}
\label{rate-et}
\gamma_{\rm et}(\omega)=\frac{2 \mu^{2}}{\hbar}\,\text{Im} \left [ \bm{n} \bar{\bm{D}}_{m}(\omega;\bm{r}_{0},\bm{r}_{0}) \bm{n}\right ]
\end{equation}
is the QE-plasmon ET rate. Some part of the transferred energy is radiated away by the  plasmonic antenna while  the rest dissipates  via the Ohmic losses in metal. These processes  are both included in the plasmon Green's function via the plasmon radiative and nonradiative rates Eqs.~(\ref{rate-rad}) and (\ref{rate-nr}), respectively, which comprise the plasmon decay rate $\gamma_{m}$ in the plasmon Green's function Eq.~(\ref{dyadic-mode}).

The QE-plasmon ET rate is obtained in explicit form by inserting the plasmon Green's function Eq.~(\ref{dyadic-mode}) into Eq.~(\ref{rate-et}),
\begin{equation}
\label{rate-mode}
\gamma_{\rm et}(\omega)
=\frac{\mu^{2}Q}{\hbar U_{m}}\, 
\frac{[\bm{n}\!\cdot\!\bm{E}(\bm{r}_{0})]^{2}}{1+4Q^{2}(\omega/\omega_{m}-1)^{2}},
\end{equation}
where $Q=\omega_{m}/\gamma_{m}$ is  the plasmon  quality factor. The  decay rate  enhancement at resonance $\omega=\omega_{m}$  is characterized by the Purcell factor  \cite{purcell-pr46}
\begin{equation}
\label{purcell-plas-gen}
F=\frac{\gamma_{\rm et}(\omega_{m})}{\gamma_{0}^{r}} = \frac{3\lambda_{m}^{3}Q}{4\pi^{2}{\cal V}_{n}}
\end{equation}
where ${\cal V}_{n}$  is the projected plasmon mode volume defined as the inverse of plasmon mode density $\rho_{n}(\bm{r})$ that characterizes the plasmon field confinement at a point $\bm{r}$ along the direction $\bm{n}$ \cite{shahbazyan-prl16,shahbazyan-prb18}:
\begin{equation}
\label{mode-volume-proj}
\dfrac{1}{{\cal V}_{n}(\bm{r})} =\rho_{n}(\bm{r})=\frac{2\left [\bm{n}\!\cdot\!\bm{E}(\bm{r})\right ]^{2}}{\int \! dV \bm{E}^{2}\,\partial (\omega_{m}\varepsilon')/\partial \omega_{m}}.
\end{equation}
Using the above expression, we find  the Purcell factor for a QE coupled to a plasmonic resonator as
\begin{equation}
\label{purcell-plas}
F
= \frac{3\lambda_{m}^{3}\,Q\,[\bm{n}\!\cdot\!\bm{E}(\bm{r}_{0})]^{2}}{2\pi^{2}\! \int \! dV \bm{E}^{2}\partial (\omega_{m}\varepsilon')/\partial \omega_{m}}.
\end{equation}
For a sufficiently large antenna, a substantial part of the energy transferred from the QE is radiated away at a rate that is proportional to the Purcell factor Eq.~(\ref{purcell-plas}). The enhancement factor for the power spectrum radiated by a QE via plasmonic antenna, relative to that  for an isolated QE, is obtained by integrating the Poynting's vector over remote surface enclosing the system \cite{shahbazyan-prb18}. The result is 
\begin{equation}
\label{enh-pl}
M(\omega)=\frac{\gamma_{\rm et}(\omega)}{\gamma_{0}^{r}}\,\eta=\frac{F\eta}{1+4Q^{2}(\omega/\omega_{m}-1)^{2}},
\end{equation}
where $\eta$ is the plasmonic antenna's radiation efficiency given by Eq.~(\ref{mode-efficiency}). The radiated power spectrum retains the  plasmon resonance  shape with maximal enhancement at resonance $M(\omega_{m})=F\eta$. Note that an increase in plasmon radiation efficiency $\eta=\gamma_{m}^{r}/(\gamma_{m}^{\rm nr}+\gamma_{m}^{r})$  also implies a reduction of the plasmon quality factor $Q=\omega_{m}/(\gamma_{m}^{\rm nr}+\gamma_{m}^{r})$ in the Purcell factor Eq.~(\ref{purcell-plas}), so that the maximal radiated power is achieved at some optimal antenna efficiency. Indeed, using Eqs.~(\ref{rate-nr}), (\ref{mode-efficiency}),  (\ref{purcell-plas-gen}) and (\ref{mode-volume-proj}), the enhancement factor at resonance is
\begin{equation}
M(\omega_{m})=F\eta=\frac{3\lambda_{m}^{3}}{4\pi^{2}}\,\frac{\zeta}{(1+\zeta)^{2}} \,\frac{\left [\bm{n}\!\cdot\!\bm{E}(\bm{r}_{0})\right ]^{2}}{\int \! dV  \varepsilon'' \bm{E}^{2}}.
\end{equation}
It is now easy to see that the maximal enhancement is achieved at  $\zeta=\gamma_{m}^{r}/\gamma_{m}^{\rm nr}=1$ corresponding to the antenna radiation efficiency $\eta=1/2$.

\section{Cooperative emission of light mediated by energy transfer to plasmonic antenna}
\label{sec:coop}

Let us now turn to  emission of light by an  \textit{ensemble} of excited QEs with dipole moments $\bm{p}_{i}$,  positioned at  $\bm{r}_{i}$ near a  plasmonic  structure. Each QE couples to the \textit{common} field generated by all QEs,
\begin{align}
\label{electric-qe}
\bm{\mathcal{E}}(\omega,\bm{r} ) = \sum_{j} \bar{\bm{D}}(\omega;\bm{r},\bm{r}_{j}) \bm{p}_{j}, 
\end{align}
where $\bar{\bm{D}}(\omega;\bm{r},\bm{r}_{j})$ is the electromagnetic Green's function Eq.~(\ref{green-nf}) in the presence of plasmonic structure. We assume that the QE's frequency is close to a particular mode frequenccy $\omega_{m}$ and only consider the \textit{weak coupling} regime, i.e., the dipole moments of individual QEs are unaffected by the coupling to a plasmon mode. Even so, the plasmon-induced correlations between QEs lead to the emergence of a collective state that transfers its energy \textit{cooperatively} to the resonant plasmon mode, while the rest of the states do not directly couple to the plasmon but can still  radiate on their own. Following the transfer,   possible energy flow pathways include plasmon reabsorption by the QEs, dissipation due to the Ohmic losses in metal, and radiation by the plasmonic antenna. In the weak coupling regime, i.e., when plasmon reabsorption is weak, and for sufficiently high antenna radiation efficiency ($\eta\sim 1$), the energy is mainly radiated away by the plasmonic antenna at about the same CET rate. Since only excited QEs participate in such one-way ET to a plasmon, the  CET rate  scales with  the number of \textit{excited} (rather total) QEs and can be tuned in a wide range by varying the excitation power \cite{strangi-prl19}. Below we present our theory for the CET-based cooperative emission.

\subsection{Plasmonic correlations, collective states and cooperative energy transfer }
 
We consider an ensemble of $N$ excited QEs with dipole moments $\bm{p}_{i}=\mu \bm{n}_{i}e^{i\phi_{i}}$, where random phases $\phi_{i}$ simulate uncorrelated initial states of QEs  after  a pulsed excitation and subsequent relaxation. Each QE interacts, via the coupling  $\bm{p}_{i}^{*} \cdot \bm{\mathcal{E}}(\bm{r}_{i})$, with the common electric field Eq.~(\ref{electric-qe}), implying that the  system eigenstates are defined by the  Green's function matrix at QEs' positions projected onto QEs' dipole moments \cite{haroche-pr82},
 \begin{equation}
 \label{coupling-matrix}
D_{ij}=\bm{p}_{i}^{*} \bar{\bm{D}}(\omega;\bm{r}_{i},\bm{r}_{j})  \bm{p}_{j}.
 \end{equation}
In the near field limit, the coupling matrix Eq.~(\ref{coupling-matrix}) can be decomposed, following the Green's function Eq.~(\ref{dyadic-mode}), into the free-space and plasmon terms, $D_{ij}=D_{ij}^{0}+D_{ij}^{m}$. The free-space coupling matrix $D_{ij}^{0}$ can, in turn, be split into direct dipole-dipole and radiative parts. The former causes random shifts of the QEs' energies which, for QEs oriented randomly, vanish \textit{on average} \cite{friedberg-pr73,shahbazyan-prb00} (we will return to this point later), while the latter has the form $D_{ij}^{0}=i(2\omega^{3}/3c^{3})\bm{p}_{i}^{*}\cdot\bm{p}_{j}$ and, in the absence of plasmonic structure, gives rise to superradiant and subradiamt states  \cite{haroche-pr82}. 

Near the plasmon resonance $\omega\sim \omega_{m}$, the coupling matrix Eq.~(\ref{coupling-matrix}) is dominated by the plasmonic term $D_{ij}^{m}$, which can also be split into non-radiative and radiative parts. The latter describes the plasmonic enhancement of radiative coupling between QEs due to resonant light scattering off the plasmonic structure, and  leads to plasmonic enhancement of Dicke supperradiance \cite{pustovit-prl09,pustovit-prb10}. Note, however, that for small plasmonic systems, the plasmon-enhanced scattering  is substantially weaker than resonant plasmon absorption \cite{bohren-book}  and, therefore, the main contribution to  the plasmon coupling matrix  comes from the non-radiative term described by the plasmon Green's function Eq.~(\ref{dyadic-mode}). Thus, for QEs' frequency close to the plasmon resonance,  the coupling matrix Eq.~(\ref{coupling-matrix})  takes the form
\begin{equation}
\label{matrix-pl}
D_{ij}^{m}(\omega)=\frac{\omega_{m}}{4 U_{m}}\frac{\bm{p}_{i}^{*}\!\cdot\!\bm{E}(\bm{r}_{i})\, \bm{p}_{j}\!\cdot\!\bm{E}(\bm{r}_{j})}{\omega_{m}-\omega-i\gamma_{m}/2}.
\end{equation}
The diagonal elements of plasmon coupling matrix Eq.~(\ref{matrix-pl}) are complex, $D_{ii}=\hbar\delta\omega^{(i)}+i\hbar \gamma_{\rm et}^{(i)}/2$, and their real and imaginary parts describe, respectively, the frequency shifts $\delta\omega^{(i)}$ and decay rates $\gamma_{\rm et}^{(i)}$  of individual QEs due to the coupling to a resonant plasmon mode.   In the weak coupling regime, the frequency shifts are relatively small and  play no significant role in the following, while  the  QE-plasmon ET  rates $\gamma_{\rm et}^{(i)}$ are given by  Eq.~(\ref{rate-mode}).

Let us now show that the plasmonic correlations between QEs, described by the non-diagonal elements of plasmon coupling matrix Eq.~(\ref{matrix-pl}), give rise to a collective state that transfers its energy \textit{cooperatively} to the  plasmon  at a rate $\gamma_{\rm et}^{N}$ that equals the sum of individual  QE-plasmon ET rates:
%
\begin{equation}
\label{rate-ens}
\gamma_{\rm et}^{N}(\omega)
=\sum_{i}\gamma_{\rm et}^{(i)}
=\frac{\mu^{2}Q}{\hbar U_{m}}
\sum_{i}\frac{[\bm{n}_{i}\!\cdot\!\bm{E}(\bm{r}_{i})]^{2}}{1+4Q^{2}(\omega/\omega_{m}-1)^{2}}.
\end{equation}
Indeed, a collective state represented by the vector $\psi_{N}=\{\bm{p}_{1}^{*}\!\cdot\!\bm{E}(\bm{r}_{1}),\dots,\bm{p}_{N}^{*}\!\cdot\!\bm{E}(\bm{r}_{N})\}$ is  an eigenstate of  the matrix Eq.~(\ref{matrix-pl}), i.e., $\hat{D}^{m}\psi_{N}=\lambda_{N}\psi_{N}$  with a complex eigenvalue  $\lambda_{N}=\hbar\delta\omega_{N}+i\hbar\gamma_{\rm et}^{N}/2$, where $\delta\omega_{N}=\sum_{i}\delta\omega^{(i)}$.
Since the imaginary part of  $\lambda_{N}$ \textit{saturates}  the ET rate from the QE ensemble to plasmon mode, the rest of the collective  states are uncoupled from that mode but, in principle, can  interact with   off-resonant modes and   radiation field. Note, however, that for large ensembles, the coupling of collective states to off-resonant modes is relatively weak \cite{vidal-prl14,petrosyan-prb17},  whereas  large plasmonic Purcell factors ensure that direct QE radiative decay is relatively weak as well,  implying that the CET to plasmonic antenna is the dominant energy flow channel. We stress that, in plasmonic systems,  the collective states emerge in response to the \textit{local} field  that can vary significantly near a plasmonic structure, rather than to the radiation field that is nearly uniform on the system scale, and, therefore, these states are distinct from the superradiant and subradiant states. Furthermore, since the QE-plasmon ET rates are determined by the spectral overlap between donors' (QEs) emission band and acceptor's (plasmon) absorption bands  \cite{novotny-book}, the CET mechanism, in contrast to superradiance \cite{haroche-pr82}, is largely insensitive to the  QEs' frequency variations  due to, e.g., direct dipole-dipole coupling between QEs, QE-metal interactions, or, in the case of semiconductor QDs, their size distribution,  as long as such variations stay within a broad plasmon resonance band.

\subsection{Cooperative Purcell factor and radiated power spectrum}

We now introduce cooperative Purcell factor for a QE ensemble  as $F_{N}=\gamma_{\rm et}^{N}(\omega_{m})/\gamma_{0}^{r}$ [compare to Eq.~(\ref{purcell-plas-gen})], where  the CET rate at  plasmon   frequency is a sum of the corresponding   individual QE-plasmon ET rates
\begin{equation}
\label{rate-cet}
\gamma_{\rm et}^{N}(\omega_{m})
=\frac{\mu^{2}Q}{\hbar U_{m}} \sum_{i} [\bm{n}_{i}\!\cdot\!\bm{E}(\bm{r}_{i})]^{2}
=\sum_{i} \frac{8\pi\mu^{2}Q}{\hbar {\cal V}_{n}^{(i)}}.
\end{equation}
Here, ${\cal V}_{n}^{(i)}$ is the  plasmon mode volume  at $\bm{r}_{i}$ projected along $\bm{n}_{i}$, given by Eq.~(\ref{mode-volume-proj}). Normalizing Eq.~(\ref{rate-cet}) by $\gamma_{0}^{r}$, we obtain cooperative Purcell factor as a sum of individual QE Purcell factors Eq.~(\ref{purcell-plas}):
\begin{equation}
\label{purcell-plas-coop0}
F_{N}=\sum_{i}\frac{3\lambda_{m}^{3}Q}{4\pi^{2}{\cal V}_{n}^{(i)}}
=\sum_{i} \frac{3\lambda_{m}^{3} Q\left |\bm{n}_{i}\!\cdot\!\bm{E}(\bm{r}_{i})\right |^{2}}{2\pi^{2}\! \int \! dV |\bm{E}|^{2}\partial (\omega_{m}\varepsilon')/\partial \omega_{m}}.
\end{equation}
Note that $F_{N}$ describes \textit{both} the plasmon  field enhancement of individual QE decay rates  characterized by  ${\cal V}_{n}^{(i)}$, and cooperative acceleration of  ensemble emission due to plasmonic correlations between QEs.


The power $W_{r}$  radiated by an ensemble of QEs coupled to plasmonic antenna is obtained in a standard way  by integrating Poynting's vector $S=(c/8\pi)|\bm{\mathcal{E}}(\omega,\bm{r})|^{2}$ over a remote surface enclosing the system \cite{novotny-book}, where  $\bm{\mathcal{E}}(\omega,\bm{r})$ is the far field generated by QEs in the presence of plasmonic structure. To extract the far field contribution from Eq.~(\ref{electric-qe}), we employ the Dyson equation for  the Green function, $\bar{\bm{D}}=\bar{\bm{D}}_{0}+\bar{\bm{D}}_{0} \chi'\bar{\bm{D}}$, where $\chi(\omega,\bm{r})$ is the plasmonic system susceptibility [compare to Eq.~(\ref{power-rad})]. Near the resonance, by replacing $\bar{\bm{D}}$ with the plasmon Green's function  Eq.~(\ref{dyadic-mode}) and using the far-field asymptotics of the free-space Green's function \cite{novotny-book}, we obtain 
\begin{equation}
\label{mode-power-rad-ens}
W_{r}= \frac{\omega^{4}}{3c^{3}} \left |\sum_{i} \left [\bm{p}_{i}+ \frac{\omega_{m}}{4 U_{m}}\frac{\bm{\mathcal{P}}\, [\bm{E}(\bm{r}_{i})\!\cdot\!\bm{p}_{i}]}{\omega_{m}-\omega-i\gamma_{m}/2}\right ] \right |^{2},
\end{equation}
where the second term describes contribution from the   antenna with dipole moment $\bm{\mathcal{P}}$. After averaging  over the random phases $\phi_{i}$ in $\bm{p}_{i}=\mu \bm{n}_{i}e^{i\phi_{i}}$ and omitting non-resonant direct QE emission (first term), we obtain the ensemble radiated power spectrum mediated by   CET to plasmonic antenna in the form
\begin{equation}
\label{mode-power-rad-ens2}
W_{r}^{N}
=\frac{\mu^{2}\omega^{4}}{3c^{3}}
\frac{\gamma_{\rm et}^{N}(\omega)}{\gamma_{0}^{r}}\,\eta
\end{equation}
where the CET rate $\gamma_{\rm et}^{N}(\omega)$ is given by Eq.~(\ref{rate-ens}) and the antenna radiation efficiency is given by Eq.~(\ref{mode-efficiency}). 
Finally, normalizing $W_{r}^{N}$ by radiated power $W_{r}^{0}=\mu^{2}\omega^{4}/3c^{3}$ of an individual QE \cite{novotny-book}, we obtain enhancement factor for radiated power spectrum as [compare to Eq.~(\ref{enh-pl})]
\begin{equation}
\label{mode-power-rad-ens3}
M_{N}(\omega)
=\frac{\gamma_{\rm et}^{N}(\omega)}{\gamma_{0}^{r}}\,\eta
=\frac{F_{N}\eta}{1+4Q^{2}(\omega/\omega_{m}-1)^{2}}.
\end{equation}
%
At plasmon resonance, the  enhancement factor is related to cooperative Purcell factor as  $M_{N}(\omega_{m})=F_{N}\eta$.

A distinct feature of CET-based cooperative emission is its robust power spectrum that retains the plasmon resonance shape with amplitude proportional to the cooperative Purcell factor, which scales with the ensemble size. In contrast to common cooperative mechanisms, e.g., superradiance, where radiation takes place directly from the QE collective states and, hence, emission spectra reflect those states' decay rates \cite{haroche-pr82}, here the light emanates from the plasmonic antenna and, hence, the emission spectra reflect the antenna's optical characteristics. For example, by placing  the plasmonic antenna  on top of a metallic mirror, the CET-based cooperative emission can be made highly directional whereas its rate can be tuned in a wide range with excitation power \cite{strangi-prl19}.  Furthermore, the radiated power spectrum  is largely insensitive to  variations of the QE  emission frequencies  as long as they fit within  a broad plasmon band; in contrast, superradiance emission spectra are strongly affected   by even  a weak disorder due to high sensitivity of subradiant states to small perturbations of the QE energies \cite{friedberg-pr73,shahbazyan-prb00}. 

\section{Discussion and numerical results}
\label{sec:num}

The CET rate Eq.~(\ref{rate-cet}) and the corresponding Purcell factor  Eq.~(\ref{purcell-plas-coop0}) depend sensitively on  the plasmon field intensity at the QEs' positions, and so we consider below two distinct cases. First, we assume that $N$ excited QEs are located in a region where the plasmon field $\bm{E}$ is approximately uniform, e.g.,  within dielectic core enclosed by a metal shell (plasmonic cavity) \cite{strangi-prl19}. In this case, after averaging over random dipole orientations in Eq.~(\ref{purcell-plas-coop0}), which results in a factor $1/3$, we obtain
\begin{equation}
F_{N}=\frac{N\lambda_{m}^{3}Q}{4\pi^{2}{\cal V}},
\end{equation}
where ${\cal V}$ is the plasmon mode volume that is related to the plasmon mode density $\rho$ as \cite{shahbazyan-prl16,shahbazyan-prb18}
\begin{equation}
\frac{1}{\cal V}=\rho(\bm{r})= \frac{2 \bm{E}^{2}(\bm{r})}{\int \! dV \bm{E}^{2}\partial (\omega_{m}\varepsilon')/\partial \omega_{m}},
\end{equation}
which, in this case, is nearly constant in the QE region. Thus, for an ensemble of $N$ excited QEs   in a region with nearly uniform plasmon field, the cooperative Purcell factor scales \textit{linearly} with $N$ on top of the plasmon field enhancement characterized by small plasmon mode volume ${\cal V}$. Correspondingly, the power radiated by the antenna increases linearly with $N$ as well, while the emission spectra retain the plasmon resonance shape  \cite{strangi-prl19}.

\begin{figure}[tb]
\begin{center}
\includegraphics[width=0.95\columnwidth]{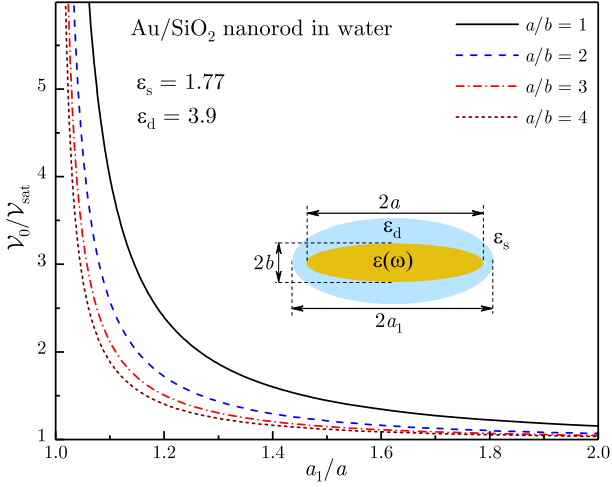}
\caption{\label{fig2} Average plasmon mode volume ${\cal V}_{0}$ in the QE region, normalized by  saturated mode volume ${\cal V}_{\rm sat}$, is plotted vs.   QE region size. Inset. Schematics of core-shell nanorod in water with Au core and QE-doped SiO$_2$ shell.}
\end{center}
\vspace{-5 mm}
\end{figure}

Consider now a common   setup, where QEs are uniformly distributed in a volume $V_{0}$ outside the metal structure, e.g., within dielectric shell on top of a metallic core (see schematics in Fig.~\ref{fig2}). In this case, the plasmon field falls off rapidly away from the metal-dielectric interface, and  so do the individual QE-plasmon ET rates. Furthermore, since the remote QEs couple weakly to  the plasmon mode,  with increasing QE region size, the cooperative Purcell factor Eq.~(\ref{purcell-plas-coop0}) should saturate.  Introducing the  \textit{average} mode volume in the QE region $V_{0}$ as \cite{shahbazyan-acsphot17}
\begin{equation}
\label{volume-mode-av}
\frac{1}{{\cal V}_{0}}
= \frac{1}{V_{0}}\!\int\! dV_{0} \,  \rho(\bm{r}) 
= \frac{1}{V_{0}}\frac{2\int \! dV_{0}\bm{E}^{2}}{\int \! dV \bm{E}^{2}\partial (\omega_{m}\varepsilon')/\partial \omega_{m}},
\end{equation}
and performing orientational averaging, we obtain the cooperative Purcell factor as
\begin{equation}
\label{purcell-plas-coop}
F_{c}=\frac{ N\lambda_{m}^{3}Q}{4\pi^{2}{\cal V}_{0}}
= \frac{n\lambda_{m}^{3}}{2\pi^{2}}\frac{Q\int \! dV_{0}\bm{E}^{2}}{\int \! dV \bm{E}^{2}\partial (\omega_{m}\varepsilon')/\partial \omega_{m}},
\end{equation}
where $n=N/V_{0}$ is the QE concentration, and subscript $c$ indicates a dependence on QE concentration rather than on  QE number. For an extended QE region saturating the plasmon mode volume, the integral  $\int \! dV_{0}\bm{E}^{2}$ is independent of  $V_{0}$ and so the average plasmon mode volume ${\cal V}_{0}$ should scale as $V_{0}$.  Using the Gauss law in Eq.~(\ref{purcell-plas-coop}) to match the electric fields at the metal-dielectric interface and omitting the remote QEs' contribution, we obtain  the saturated plasmon mode volume ${\cal V}_{\rm sat}$ as,
\begin{equation}
\label{mode-volume-sat}
\frac{{\cal V}_{\rm sat}}{V_{0}} 
=\frac{\varepsilon_{d}\,\omega_{m} }{2|\varepsilon'(\omega_{m})| }   \frac{\partial \varepsilon'(\omega_{m}) }{\partial \omega_{m} },
\end{equation}
where $\varepsilon_{d}$ is the dielectric constant of QE region. Finally, for QEs saturating the plasmon mode volume, we obtain the cooperative Purcell factor,
\begin{equation}
\label{purcell-plas-coop-sat}
F_{\rm sat}=\frac{N\lambda_{m}^{3}Q}{4\pi^{2}{\cal V}_{\rm sat}}
= 
\frac{ n\lambda_{m}^{3}}{2\pi^{2} \varepsilon_{d}}
\,\frac{Q|\varepsilon'(\omega_{m})|}{  \omega_{m}\partial \varepsilon'(\omega_{m})/\partial \omega_{m}},
\end{equation}
which scales linearly with the QE concentration $n$,  and is independent of local fields. Note that, in the saturated case,  the cooperative Purcell factor depends on the system geometry only via  the plasmon  frequency $\omega_{m}$. With increasing  QE region size, the radiated power spectrum Eq.~(\ref{mode-power-rad-ens3}) saturates as well,
\begin{equation}
\label{enh-plas-coop-sat}
M_{\rm sat}=F_{\rm sat} \eta  
= 
\frac{ n\lambda_{m}^{3}}{4\pi^{2}\varepsilon_{d}}\left| \frac{\varepsilon'(\omega_{m})}{\varepsilon''(\omega_{m})}\right |\frac{\zeta}{(1+\zeta)^{2}},
\end{equation}
where $\zeta$ is given by Eq.~(\ref{mode-rate-relative}) and we used Eq.~(\ref{rate-nr}). The maximal enhancement is achieved at $\zeta=1$, corresponding to the antenna radiation efficiency $\eta=1/2$. Note that, in the saturated regime, the radiated power is determined, apart from a large factor $|\varepsilon'(\omega_{m})/\varepsilon''(\omega_{m})|\gg 1$, by the QE number that would fit, at a given QE concentration $n$, within the volume $\lambda_{m}^{3}$, independent of the system size. Mode volume saturation was recently observed in photoluminescence of strongly coupled molecular excitons excited in J-aggregates embedded within dielectric shell enclosing a gold nanoprism \cite{shegai-nl17}.

%
%

As an example, here we present the results of numerical calculations for core-shell nanorods in water with  QEs embedded, at constant concentration $n$, within SiO$_{2}$ shell  enclosing an Au core [see schematics in Fig.~\ref{fig2}].  We model  this core-shell structure by two confocal prolate spheroids with semimajor axes $a$ and $a_{1}$, corresponding to Au/SiO$_2$ and SiO$_2$/H$_2$O interfaces, respectively. The QEs' emission frequency was kept in resonance with the longitudinal dipole plasmon mode, and  the experimental Au dielectric function was used in all calculations. 

\begin{figure}[tb]
\begin{center}
\includegraphics[width=0.95\columnwidth]{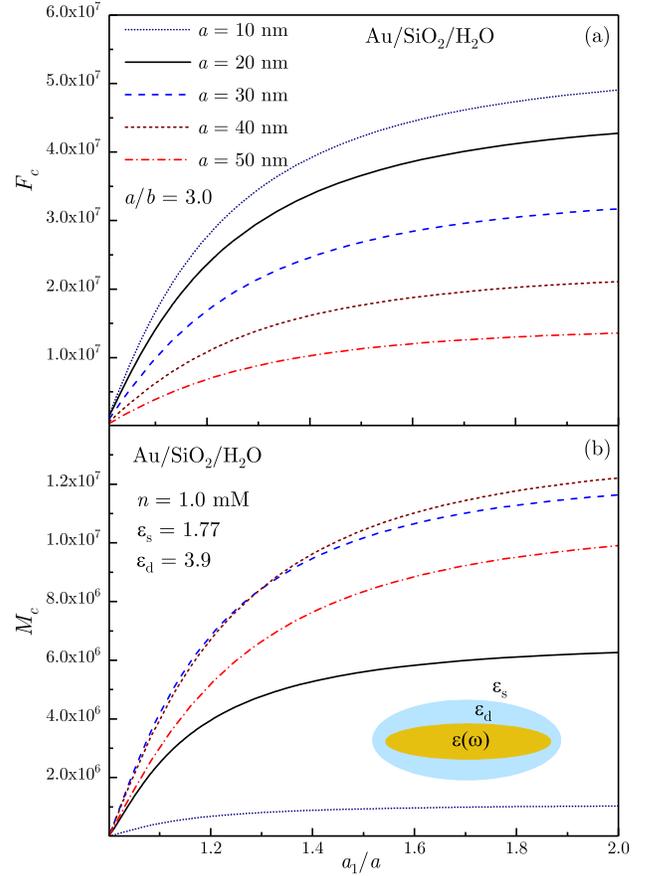}
\caption{\label{fig3}  
Cooperative Purcell factor (a) and power spectrum enhancement factor  (b) are plotted vs. QE region size for several  nanorod lengths  at fixed core aspect ratio $a/b=3.0$.}
\end{center}
\vspace{-5 mm}
\end{figure}
%

In  Fig.~\ref{fig2}, we plot the average plasmon mode volume ${\cal V}_{0}$ within SiO$_2$ shell, normalized by saturated mode volume ${\cal V}_{\rm sat}$ (i.e., for infinitely thick shell), versus $a_{1}/a$. Note that, with expanding QE region, the average plasmon mode volume  in that region takes an increasingly smaller fraction of the  region  volume $V_{0}$ until it saturates to the value  Eq.~(\ref{mode-volume-sat}). While for thin shells, the ratio ${\cal V}_{0}/{\cal V}_{\rm sat}$  is relatively large due to significant field spillover beyond the QE region, it rapidly saturates with increasing $a_{1}/a$, reaching the value ${\cal V}_{0}/{\cal V}_{\rm sat}\approx 1$ at $a_{1}/a\approx 2$.  

In Fig.~\ref{fig3}, we  plot calculated cooperative Purcell factor $F_{c}$ and power spectrum enhancement factor  $M_{c}=F_{c}\eta$ versus QE region size for several nanorod lengths $a$ at fixed aspect ratio $a/b=3.0$. With expanding QE region, the initial increase of $F_{c}$ and $M_{c}$ saturates as the remote QEs' coupling to plasmon weakens. Although the plasmon mode volume is nearly saturated at $a_{1}/a=2.0$ (see Fig.~\ref{fig2}), both  $F_{c}$ and $M_{c}$ still change, albeit  weakly, with the QE region size  due to  plasmon frequency dependence on the shell thickness. The magnitudes of $F_{c}$ and $M_{c}$ for different nanorod lengths are determined by the interplay between the plasmon quality factor $Q$ and antenna radiation efficiency $\eta$, which change in the opposite way as  the plasmon radiative decay rate $\gamma_{m}^{r}$ increases. In fact, the smallest nanorod ($a=10$ nm) has the  largest Purcell factor $F_{c}\propto Q$ and the smallest enhancement factor $M_{c}\propto Q\eta$, while the   largest enhancement factors are achieved for  medium-sized rods. Note that, according to Eq.~(\ref{enh-plas-coop-sat}), the optimal antenna radiation efficiency  for maximal power spectrum enhancement  is $\eta=1/2$.

We now turn to comparison between cooperative acceleration of the ensemble emission rate and  plasmon field enhancement of the individual QE at a hot spot. Although the CET mechanism incorporates  the LDOS enhancement as well,   randomly distributed QEs can miss the hot spot, characterized by extremely large LDOS, and so, at  low QE concentration, the LDOS enhancement of the QE decay rate  at a hot spot can be comparable to cooperative acceleration of the ensemble decay rate. To illustrate this point, let as compare the individual QE-plasmon ET rate for a  QE  near sharp tip of   small metal nanostructure to the  CET rate for a  QE ensemble  saturating the plasmon mode volume near the same structure. The Purcel factor near  a metal tip has the form \cite{shahbazyan-prb18}
\begin{equation}
\label{purcell-plas-tip}
F_{\rm tip}=\frac{3\lambda_{m}^{3}Q}{2\pi^{2}V_{\rm met}}\,
\frac{|\varepsilon'(\omega_{m})|^{2}}
{\omega_{m}\partial \varepsilon'(\omega_{m})/\partial \omega_{m}} \,
\bigl[ \bm{n}\!\cdot\!\tilde{\bm{E}} (\bm{r}_{0})\bigr]^{2},
\end{equation}
where $\tilde{\bm{E}}(\bm{r}_{0})$ is the plasmon  field   at point $\bm{r}_{0}$  normalized by its value at the  metal-dielectric  interface, $\bm{n}$ is the QE dipole orientation, and $V_{\rm met}$ is the metal  volume. Using Eq.~(\ref{purcell-plas-coop-sat}), the ratio of Purcell factors takes the form
\begin{equation}
\label{purcell-ratio}
\frac{F_{\rm sat}}{F_{\rm tip}}=\frac{nV_{\rm met}}{3\varepsilon_{d}|\varepsilon'(\omega_{m})|
\bigl[ \bm{n}\!\cdot\!\tilde{\bm{E}}(\bm{r}_{0})\bigr]^{2}}.
\end{equation}
Although the normalized field $\tilde{\bm{E}}$ reaches unity at the (classical)  interface, its magnitude is, in fact, significantly damped due to  nonlocal effects near the metal surface \cite{smith-science12,mortensen-nc14}, so that  $\tilde{\bm{E}}^{2}< 1$ even at a hot spot. 


%
\begin{figure}[tb]
\begin{center}
\includegraphics[width=0.95\columnwidth]{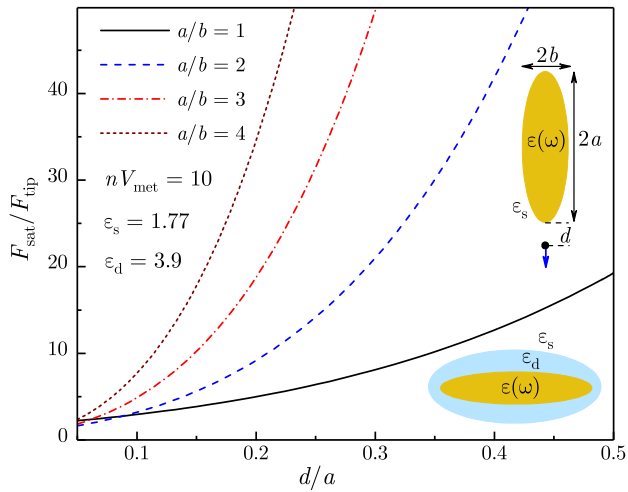}
\caption{\label{fig4}  
Cooperative acceleration of the ensemble decay rate relative to LDOS enhancement of the individual QE  decay rate  is plotted vs. a QE distance $d$ to the metal tip. Lower inset: Schematics of a Au nanorod surrounded by QE-doped SiO$_2$ shell in water. Upper inset: Schematics of a QE located at a distance $d$ from the Au nanorod tip in water.}
\end{center}
\vspace{-5 mm}
\end{figure}
%

In  Fig.~\ref{fig4},  we compare  the cooperative Purcell factor Eq.~(\ref{purcell-plas-coop-sat})  for an ensemble of QEs uniformly distributed within dielectric shell around an Au nanorod to the individual Purcel factor Eq.~(\ref{purcell-plas-tip}) for a QE placed near same Au nanorod tip (see schematics in Fig.~\ref{fig4}). Since  the plasmon mode volume near a metal tip scales with the metal volume $V_{\rm met}$ \cite{shahbazyan-prb18}, the ratio $F_{\rm sat}/F_{\rm tip}$ scales as $nV_{\rm met}$, while being very sensitive to  QE's  distance $d$ to the tip  [see Eq.~(\ref{purcell-ratio})]. For a relatively low value $nV_{\rm met}=10$, i.e., merely ten QEs that would fit within metal volume at given QE concentration $n$ in the shell, the  ensemble CET decay rate is comparable to individual QE decay rate at a hot spot, but gets much larger once the QE is moved away from the tip.  Note that, for $d< 1$ nm, the local fields are damped by nonlocal effects \cite{smith-science12,mortensen-nc14}. For higher QE concentrations,   cooperative acceleration of the   emission rate is well beyond  the field enhancement limit \cite{strangi-prl19}.
%


\section{Conclusions}
\label{sec:conc}

Finally, let us discuss an important distinction  between CET-based cooperative emission and traditional cooperative emission, such as superfluorescence. When an ensemble of QEs confined within some small volume is excited by a short pulse, the system undergoes  coherence build-up, which involves multiple processes of photon emission and resbsorption until all QE dipoles get aligned, followed by superradiant burst \cite{haroche-pr82}. The collective superradiant state emerges as a result of QEs’ interaction with their common radiation field, and the radiation takes place directly from QEs that form a superradiant state. 
In contrast, when QEs are placed near a  plasmonic resonator, they inreract strongly, via near-field coupling, with the plasmon mode, while direct radiative coupling between QEs is much weaker. Therefore, the collective states are formed due to emitters' correlations via the plasmon field, which can vary significantly depending on system geometry, rather than the radiation field, which is nearly uniform on the system scale. Accordingly, the collective states participating in CET-based cooperative emission are completely different from the traditional superradiant and subradiant states. Importantly, in CET-based cooperative emission, the light emanates from the antenna  and, therefore, the emission spectrum retains the lineshape of plasmon resonance, in contrast to superradiance spectra, which reflects the decay rates of superradiant and subradiant collective states. Furthermore, in the weak coupling regime, the plasmon reabsorption is suppressed and, hence, the CET-based emission does not imply coherence build-up and, therefore, in contrast to superfluorescence, does not take place through a burst, but instead represents a steady process occurring at rate controlled by the  excitation power \cite{strangi-prl19}.

In summary, we developed a theory of cooperative emission from an ensemble of QEs mediated by CET  to a plasmonic antenna at a rate equal to the sum of individual QE-plasmon ET rates. If the antenna radiation efficiency is sufficiently high, the transferred energy is radiated away at approximately the same cooperative rate that scales with the ensemble size. We obtained explicit expressions, in terms of local fields, for cooperative Purcell factor and enhancement factor for radiated power spectrum valid for plasmonic nanostructures of any shape with characteristic size smaller than the radiation wavelength. The radiated power spectrum retains the plasmon resonance lineshape with overall amplitude scaling with the ensemble size. We discussed typical experimental geometries, where (a)  QEs are located inside a plasmonic nanocavity with nearly uniform plasmon LDOS, and (b) QEs are distributed in a region outside a metal nanostructure characterized by highly nonuniform LDOS. In the former case,  the CET-based cooperative emission rate scales linearly with the number of excited QEs  \cite{strangi-prl19}, which stands  in stark contrast to superradiant emission   at a rate that scales with the total QE number even if a few QEs are excited. In the latter case, we show that for extended QE regions saturating the plasmon mode volume, the ensemble emission rate and, correspondingly, cooperative Purcell factor have universal form independent of the  system size.   We have also compared cooperative acceleration of the ensemble decay rate to the plasmon LDOS enhancement of an individual QE decay rate  at a hot spot near metallic tip. Although the CET mechanism does incorporate the plasmon LDOS enhancement as well, the  QE emission rate  at a hot spot can be comparable to the CET rate  since, for small QE ensembles,   randomly distributed QEs can easily miss the hot spot. Finally, we have shown analytically and numerically that the strongest enhancement of radiated power  does not follow the largest Purcell factor but, instead, is achieved for structures with the optimal antenna radiation efficiency $\eta=1/2$.

\acknowledgments

This work was supported in part  by the National Science Foundation under Grants No. DMR-1610427,  No. DMR-1826886, and No. HRD-1547754.





\end{document}